\documentclass[11pt]{article}
\usepackage{graphicx}
\usepackage{tabularx}
\usepackage{url}
\usepackage{amsthm}
\usepackage{amsmath}
\usepackage{amsfonts}
\usepackage[labelfont=bf]{caption}

\theoremstyle{plain}
\newtheorem{theorem}{Theorem}
\newtheorem{lemma}{Lemma}

\theoremstyle{definition}
\newtheorem{definition}{Definition}

\newtheorem{example}{Example}
\theoremstyle{remark}

\date{}

\begin{document}

\title{Deterministic and Strongly Nondeterministic Decision Trees for
Decision Tables from Closed Classes}

\author{Azimkhon Ostonov and Mikhail Moshkov \\
Computer, Electrical and Mathematical Sciences \& Engineering Division \\ and Computational Bioscience Research Center\\
King Abdullah University of Science and Technology (KAUST) \\
Thuwal 23955-6900, Saudi Arabia\\ \{azimkhon.ostonov,mikhail.moshkov\}@kaust.edu.sa
}

\maketitle

\begin{abstract}
In this paper, we consider classes of decision tables with 0-1-decisions closed relative to
removal of attributes (columns) and changing decisions assigned to rows. For
tables from an arbitrary closed class, we study the dependence of the
minimum complexity of deterministic decision trees on various parameters of the
tables: the minimum complexity of a test, the complexity of the set of attributes attached to columns,
and the minimum complexity of a strongly nondeterministic decision tree. We also study the dependence of the
minimum complexity of strongly nondeterministic decision trees on the complexity of the set of attributes attached to columns. Note that a strongly nondeterministic decision tree can be interpreted as a set of true decision rules that cover all rows labeled with the decision $1$.
\end{abstract}

{\it Keywords}: closed classes of decision tables, deterministic decision trees, strongly nondeterministic decision trees

\section{Introduction\label{S0}}

In this paper, we consider closed classes of decision tables with
0-1-decisions and study relationships among four parameters of
these tables: the minimum complexity of a deterministic decision tree,
the minimum complexity of a strongly nondeterministic decision tree, the complexity of the set of attributes attached to columns, and the minimum complexity of a test -- a set of attributes that separate any two rows with different decisions.

A decision table with 0-1-decisions is a rectangular table in which
columns are labeled with attributes, rows are pairwise different and each
row is labeled with a decision from the set $\{0,1\}$. Rows are interpreted
as tuples of values of the attributes. For a given row, it is required to
find the decision attached to the row. To this end,
we can use the following queries: we can choose an attribute and ask what is
the value of this attribute in the considered row. We study two types of
algorithms based on these queries: deterministic and strongly nondeterministic
decision trees. One can interpret strongly nondeterministic decision trees for a
decision table as a way to represent an arbitrary system of true decision
rules for this table that cover all rows labeled with the decision $1$. We consider a fairly wide class of complexity measures that characterize the time complexity of decision
trees. Among them, we distinguish so-called bounded complexity measures, for
example, the depth of decision trees.

Conventional decision tables including decision tables with 0-1-decisions appear in data analysis and in such areas as combinatorial optimization, computational geometry, and fault diagnosis, where they are used to represent and explore problems \cite{Moshkov05,MoshkovZ11}.

Decision trees \cite%
{AbouEishaACHM19,book20,BreimanFOS84,Moshkov05,Moshkov20,Quinlan93,RokachM07}
and decision rule systems  \cite%
{BorosHIK97,BorosHIKMM00,ChikalovLLMNSZ13,FurnkranzGL12,MPZ08,MoshkovZ11,Pawlak91,PawlakS07}
are widely used as classifiers, as a means for knowledge representation, and
as algorithms for solving various problems of combinatorial optimization,
fault diagnosis, etc. Decision trees and rules are among the most
interpretable models in data analysis \cite{Molnar22}. Strongly nondeterministic decision trees (systems of true decision rules covering all rows labeled with the decision $1$) for decision tables with 0-1-decisions are less known.
Note that the depth of  strongly nondeterministic decision trees for computation Boolean functions (variables of a function are considered as attributes) was studied in \cite{Moshkov95}. Strongly nondeterministic decision trees for $(2)$-problems over information systems (see the next paragraph for the explanations) were studied  in the book \cite{Moshkov20}.

We study classes of decision tables with 0-1-decisions closed under
two operations: removal of columns and changing of decisions. The most natural examples of such
classes are closed classes of decision tables generated by information
systems \cite{Pawlak81}. An information system consists of a set of objects
(universe) and a set of attributes (functions) defined on the universe and
with values from a finite set. A $(2)$-problem over an information system is
specified by a finite number of attributes that divide the universe into
nonempty domains in which these attributes have fixed values. A decisions from the set $\{0,1\}$ is attached to each domain. For a given object from
the universe, it is required to find the decision attached to the
domain containing this object.

A decision table with 0-1-decisions
corresponds to this problem in a natural way: columns of this table are
labeled with the considered attributes, rows correspond to domains and are
labeled with decisions attached to domains. The set of decision
tables corresponding to $(2)$-problems over an information system forms a closed class
generated by this system. Note that the family of all closed classes is
essentially wider than the family of closed classes generated by information
systems. In particular, the union of two closed classes generated by two
information systems is a closed class. However, generally, there is no an
information system that generates this class.

Various classes of objects that are closed under different operations are
intensively studied. Among them, in particular, are classes of Boolean
functions closed under the operation of superposition \cite{Post41} and
minor-closed classes of graphs \cite{Robertson04}. Decision tables with 0-1-decisions represent an interesting mathematical
object deserving mathematical research, in particular, the study of closed
classes of such decision tables.

This paper continues the study of closed classes of decision tables that began
with work \cite{Moshkov89} and continued with work \cite{Ostonov23}. In
\cite{Moshkov89}, we studied the dependence of the minimum depth of
deterministic decision trees and the depth of deterministic decision trees
constructed by a greedy algorithm on the number of attributes (columns) for
conventional decision tables from classes closed under operations of removal
of columns and changing of decisions.

In \cite{Ostonov23}, we considered classes of decision tables
with many-valued decisions closed under operations of removal of columns,
changing of decisions, permutation of columns, and duplication of columns.
We studied relationships among three parameters of these tables: the
complexity of a decision table (if we consider the depth of decision trees,
then the complexity of a decision table is the number of columns in it), the
minimum complexity of a deterministic decision tree, and the minimum
complexity of a nondeterministic decision tree. We considered rough
classification of functions characterizing relationships and enumerated all possible seven types of the relationships.

In this paper, we study four functions:  $F_{\psi ,A}^{W}(n)$, $F_{\psi ,A}^{\Theta }(n)$, $F_{\psi ,A}(n)$, and $G_{\psi ,A}(n)$.

The function $F_{\psi ,A}^{W}(n)$ characterizes the growth in the worst case of the minimum complexity of a deterministic decision tree for a decision table with the growth of the complexity of the set of attributes attached to columns of the table.
The function $F_{\psi ,A}^{\Theta }(n)$ characterizes the growth in the worst case of the minimum complexity of a deterministic decision tree for a decision table with the growth of the minimum complexity of a test of the table.  In particular, we proved that the functions $F_{\psi ,A}^{W}(n)$ and  $F_{\psi ,A}^{\Theta }(n)$ are either bounded from above by a constant, or grow as a logarithm of $n$, or grow almost linearly depending on $n$ (they are bounded from above by $n$ and are equal to $n$ for infinitely many $n$).

The function $F_{\psi ,A}(n)$ characterizes the growth in the worst case of the minimum complexity of a deterministic decision tree for a decision table with the growth of the minimum complexity of a strongly nondeterministic decision tree for the table. We indicated the condition for the function $F_{\psi ,A}(n)$ to be defined
everywhere. Let $F_{\psi ,A}(n)$ be everywhere defined. We proved that this function is either bounded from above by a constant, or is greater than or equal to $n$ for infinitely many $n$. In particular, the function $F_{\psi ,A}(n)$ can grow as an arbitrary  nondecreasing function  $\varphi$ such that $\varphi (n)\geq n$ and $\varphi (0)=0$. We indicated also conditions for the function $F_{\psi ,A}(n)$ to be bounded from above by a polynomial on $n$.

The function $G_{\psi ,A}(n)$ characterizes the growth in the worst case of the minimum complexity of a strongly nondeterministic decision tree for a decision table with the growth of the complexity of the set of attributes attached to columns of the table. In particular, we proved that the function $G_{\psi ,A}(n)$ is either bounded from above by a constant or grows almost linearly depending on $n$ (it is bounded from above by $n$ and is equal to $n$ for infinitely many $n$).

There is a similarity between some results obtained in this paper for closed classes of decision tables and  results from the book \cite{Moshkov20} obtained for $(2)$-problems over information systems. However, the results obtained in the present paper are more general.

The paper consists of eight sections. In Sect. \ref{S1}, main definitions and notation are considered. In Sect. \ref{S2}, we provide the main results. Section \ref{S3} contains auxiliary statements. In Sects. \ref{S4}-\ref{S6}, we prove the main results. Section \ref{S7} contains short conclusions.

\section{Main Definitions and Notation\label{S1}}

Denote $\omega =\{0,1,2,\ldots \}$ and, for any $k\in \omega \setminus
\{0,1\}$, denote $E_{k}=\{0,1,\ldots ,k-1\}$. Let $P=\{f_{i}:i\in \omega \}$
be the set of \emph{attributes} (really names of attributes). Two attributes $f_i,f_j \in P$ are considered different if $i \neq j$.

\subsection{Decision Tables\label{S1.1}}

First, we define the notion of a decision table with 0-1-decisions.
\begin{definition}
    Let $k\in \omega \setminus \{0,1\}$. Denote by $\mathcal{M}_{k}^{2}$ the set
of rectangular tables filled with numbers from $E_{k}$ in each of which rows
are pairwise different, each row is labeled with a number from $E_{2}$
(decision), and columns are labeled with pairwise different attributes from  $P$. Rows are interpreted as tuples of values of these attributes. Empty tables without rows belong also to the set $\mathcal{M}_{k}^{2}$. We will use the same notation $\Lambda$ for these tables. Tables from $\mathcal{M}_{k}^{2}$ will be
called \emph{decision tables with 0-1-decisions} or simply \emph{decision tables}. 
\end{definition}
Two tables from $\mathcal{M}_{k}^{2}$ are
\emph{equal} if one can be obtained from another by permutation of rows with
attached to them decisions.
\begin{example}
Figure \ref{fig1} shows a decision table from $\mathcal{M}_{k}^{2}$.
\end{example}

\begin{figure}
\begin{minipage}[c]{1.0\textwidth}
\begin{center}
\begin{tabular}{ |ccc|c| }
 \hline
 $f_{2}$ & $f_{4}$ & $f_{3}$ &\\
  \hline
 1 & 1 & 1 & $0$ \\
 0 & 1 & 1 & $0$ \\
 1 & 1 & 0 & $1$ \\
 0 & 0 & 1 & $1$ \\
 1 & 0 & 0 & $1$ \\
 0 & 0 & 0 & $1$ \\
 \hline
\end{tabular}
\end{center}
\end{minipage}
\caption{Decision table from $\mathcal{M}_{2}^{2}$}
\label{fig1}
\end{figure}

Denote by $\mathcal{M}_{k}^{2}\mathcal{C}$ the set of tables from $\mathcal{M%
}_{k}^{2}$ in each of which all rows are labeled with the same decision. Let
$\Lambda \in $ $\mathcal{M}_{k}^{2}\mathcal{C}$.

Let $T$ be a nonempty table from $\mathcal{M}_{k}^{2}$. Denote by $P(T)$ the
set of attributes attached to columns of the table $T$. Let $%
f_{i_{1}},\ldots ,f_{i_{m}}\in P(T)$ and $\delta _{1},\ldots ,\delta _{m}\in
E_{k}$. We denote by $T(f_{i_{1}},\delta _{1})\cdots (f_{i_{m}},\delta _{m})$
the table obtained from $T$ by removal of all rows that do not satisfy the
following condition: in columns labeled with attributes $f_{i_{1}},\ldots
,f_{i_{m}}$, the row has numbers $\delta _{1},\ldots ,\delta _{m}$,
respectively.

We now define two operations on decision tables: removal of columns and changing of decisions. Let $T\in \mathcal{M}_{k}^{2}$.

\begin{definition}
\emph{Removal of columns}.
Let $D\subseteq P(T)$. We remove from $T$ all
columns labeled with the attributes from the set $D$. In each group of rows
equal on the remaining columns, we keep one with the minimum decision.
Denote the obtained table by $I(D,T)$. In particular, $I(\emptyset ,T)=T$
and $I(P(T),T)=\Lambda $. It is obvious that $I(D,T)\in $ $\mathcal{M}%
_{k}^{2}$.
\end{definition}

\begin{definition}
\emph{Changing of decisions}.
Let $\nu :E_{k}^{\left\vert P(T)\right\vert }\rightarrow E_{2}$ (by
definition, $E_{k}^{0}=\emptyset $). For each row $\bar{\delta}$ of the
table $T$, we replace the decision attached to this row with $\nu (\bar{%
\delta})$. We denote the obtained table by $J(\nu ,T)$. It is obvious that $%
J(\nu ,T)\in $ $\mathcal{M}_{k}^{2}$.
\end{definition}

\begin{definition}
Denote $\left[ T\right] =\{J(\nu ,I(D,T)):D\subseteq P(T),\nu
:E_{k}^{\left\vert P(T)\setminus D\right\vert }\rightarrow E_{2}\}$. The set $\left[ T\right]$ is the \emph{closure of the table} $T$ under the operations of removal of columns and changing of decisions.
\end{definition}
\begin{example}
Figure \ref{fig2} shows the table $J(\nu ,I(D,T))$, where $T$ is the table
shown in Fig. \ref{fig1}, $D=\{f_{4}\}$ and $\nu (x_{1},x_{2})=x_{1}\vee x_{2}$.
\end{example}

\begin{figure}
\begin{minipage}[c]{1.0\textwidth}
\begin{center}
\begin{tabular}{ |cc|c| }
 \hline
 $f_{2}$ & $f_{3}$ &\\
  \hline
 1 & 1 & $1$ \\
 0 & 1 & $1$ \\
 1 & 0 & $1$ \\
 0 & 0 & $0$ \\
 \hline
\end{tabular}
\end{center}
\end{minipage}
\caption{Decision table obtained from the decision table shown in Fig. \ref{fig1} by removal of a column and changing of decisions}
\label{fig2}
\end{figure}

\begin{definition}
Let $A\subseteq \mathcal{M}_{k}^{2}$ and $A\neq \emptyset $. Denote $\left[ A%
\right] =\bigcup_{T\in A}\left[ T\right] $. The set $\left[ A\right]$ is the \emph{closure of the set} $A$ under the considered two operations. The class (the set) of decision
tables $A$ will be called a \emph{closed class} if $\left[ A\right] =A$.
\end{definition}
Let $A_1$ and $A_2$ be closed classes of decision tables from $\mathcal{M}_{k}^{2}$. Then $A_1 \cup A_2$ is a closed class of decision tables from $\mathcal{M}_{k}^{2}$.

\subsection{Tests\label{S1.2}}
We now define the notion of a test for a decision table.
\begin{definition}
    Let $T\in $ $\mathcal{M}_{k}^{2}$. A set of attributes $D\subseteq P(T)$
will be called a \emph{test} of the decision table $T$ if in columns labeled with attributes from $D$ any two rows of $T$ with
different decisions are different. By definition, for any table $T \in \mathcal{M}_{k}^{2}\mathcal{C}$, any subset of the set  $P(T)$ including the empty set  is a test of $T$.
\end{definition}
\begin{example}
The set $\{f_{4},f_{3}\}$ is a test of the decision table
shown in Fig. \ref{fig1}.
\end{example}

\subsection{Deterministic and Strongly Nondeterministic Decision Trees\label%
{S1.3}}

A \emph{finite tree with root} is a finite directed tree in which exactly
one node called the \emph{root} has no entering edges. The nodes without
leaving edges are called \emph{terminal} nodes.

\begin{figure}
    \centering
    \includegraphics[width=0.305\columnwidth]{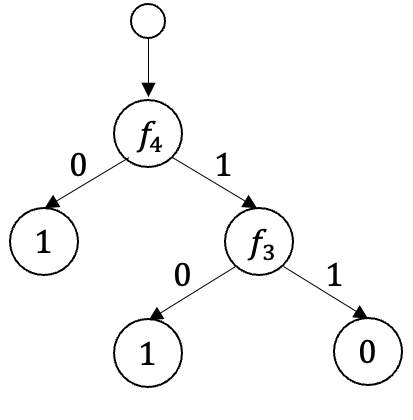}
    \caption{A deterministic decision tree for the decision table shown in Fig. \ref{fig1}}
    \label{fig3}
\end{figure}

\begin{figure}
    \centering
    \includegraphics[width=0.245\columnwidth]{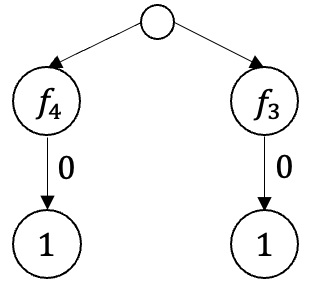}
    \caption{A strongly nondeterministic decision tree for the decision table shown in Fig. \ref{fig1}}
    \label{fig4}
\end{figure}

\begin{definition}
A $k$-\emph{decision tree} is a finite tree with root, which has at least
two nodes and in which

\begin{itemize}
\item The root and edges leaving the root are not labeled.

\item Each terminal node is labeled with a decision from the set $E_{2}$.

\item Each node, which is neither the root nor a terminal node, is labeled
with an attribute from the set $P$. Each edge leaving such node is labeled
with a number from the set $E_{k}$.
\end{itemize}
\end{definition}
\begin{example}
Figures \ref{fig3} and \ref{fig4} show $2$-decision trees.
\end{example}

We denote by $\mathcal{C}_{k}$ the set of all $k$-decision trees. Let $%
\Gamma \in \mathcal{C}_{k}$. We denote by $P(\Gamma )$ the set of attributes
attached to nodes of $\Gamma $ that are neither the root nor terminal nodes.
A \emph{complete path} of $\Gamma $ is a sequence $\tau =v_{1},d_{1},\ldots
,v_{m},d_{m},v_{m+1}$ of nodes and edges of $\Gamma $ in which $v_{1}$ is
the root of $\Gamma $, $v_{m+1}$ is a terminal node of $\Gamma $ and, for $%
j=1,\ldots ,m$, the edge $d_{j}$ leaves the node $v_{j}$ and enters the node
$v_{j+1}$. Let $T\in $ $\mathcal{M}_{k}^{2}$. If $P(\Gamma )\subseteq P(T)$,
then we correspond to the table $T$ and the complete path $\tau $ a decision
table $T(\tau )$. If $m=1$, then $T(\tau )=T$. If $m>1$ and, for $j=2,\ldots
,m$, the node $v_{j}$ is labeled with the attribute $f_{i_{j}}$ and the edge
$d_{j}$ is labeled with the number $\delta _{j}$, then $T(\tau
)=T(f_{i_{2}},\delta _{2})\cdots (f_{i_{m}},\delta _{m})$.

\begin{definition}
Let $T\in $ $\mathcal{M}_{k}^{2}\setminus \{\Lambda \}$. A \emph{%
deterministic decision tree} for the table $T$ is a $k$-decision tree $%
\Gamma $ satisfying the following conditions:

\begin{itemize}
\item Only one edge leaves the root of $\Gamma $.

\item For any node, which is neither the root nor a terminal node, edges
leaving this node are labeled with pairwise different numbers.

\item $P(\Gamma )\subseteq P(T)$.

\item For any row of $T$, there exists a complete path $\tau $ of $\Gamma $
such that the considered row belongs to the table $T(\tau )$.

\item For any complete path $\tau $ of $\Gamma $, either $T(\tau )=\Lambda $
or all rows of $T(\tau )$ are labeled with the decision attached to the
terminal node of $\tau $.
\end{itemize}
\end{definition}
\begin{example}
The $2$-decision tree shown in Fig. \ref{fig3} is a deterministic
decision tree for the decision table shown in Fig. \ref{fig1}.
\end{example}

\begin{definition}
    Let $T\in $ $\mathcal{M}_{k}^{2}\setminus \mathcal{M}_{k}^{2}\mathcal{C}$. A
\emph{strongly nondeterministic decision tree} for the table $T$ is a $k$%
-decision tree $\Gamma $ satisfying the following conditions:

\begin{itemize}
\item Each terminal node is labeled with the decision $1$.

\item $P(\Gamma )\subseteq P(T)$.

\item For any row of $T$ labeled with the decision $1$, there exists a
complete path $\tau $ of $\Gamma $ such that the considered row belongs to
the table $T(\tau )$.

\item For any complete path $\tau $ of $\Gamma $, either $T(\tau )=\Lambda $
or all rows of $T(\tau )$ are labeled with the decision $1$.
\end{itemize}
\end{definition}
\begin{example}
The $2$-decision tree shown in Fig. \ref{fig4} is a strongly
nondeterministic decision tree for the decision table shown in Fig. \ref{fig1}.
\end{example}

\subsection{Complexity Measures\label{S1.4}}

Denote by $B$ the set of all finite words over the alphabet $P=\{f_{i}:i\in
\omega \}$, which contains the empty word $\lambda $ and on which the word
concatenation operation is defined.

\begin{definition}
    A \emph{complexity measure} is an arbitrary function $\psi :B\rightarrow \omega $
that has the following properties: for any words $\alpha _{1},\alpha _{2}\in
B$,

\begin{itemize}
\item $\psi (\alpha _{1})=0$ if and only if $\alpha _{1}=\lambda $ -- \emph{%
positivity} property.

\item $\psi (\alpha _{1})=\psi (\alpha _{1}')$ for any 
word $\alpha _{1}'$ obtained from $\alpha _{1}$ by permutation of letters  -- \emph{%
commutativity} property.

\item $\psi (\alpha _{1})\leq \psi (\alpha _{1}\alpha _{2})$ -- \emph{%
nondecreasing} property.

\item $\psi (\alpha _{1}\alpha _{2})\leq \psi (\alpha _{1})+\psi (\alpha
_{2})$ -- \emph{boundedness from above} property.
\end{itemize}
\end{definition}

The following functions are complexity measures:

\begin{itemize}
\item Function $h$ for which, for any word $\alpha \in B$, $h(\alpha
)=\left\vert \alpha \right\vert $, where $\left\vert \alpha \right\vert $ is
the length of the word $\alpha $.

\item An arbitrary function $\varphi :B\rightarrow \omega $ such that $%
\varphi (\lambda )=0$, for any $f_{i}\in B$, $\varphi (f_{i})>0$ and, for
any nonempty word $f_{i_{1}}\cdots f_{i_{m}}\in B$,
\begin{equation}
\varphi (f_{i_{1}}\cdots f_{i_{m}})=\sum_{j=1}^{m}\varphi (f_{i_{j}}).
\label{E1}
\end{equation}

\item An arbitrary function $\rho :B\rightarrow \omega $ such that $\rho
(\lambda )=0$, for any $f_{i}\in B$, $\rho (f_{i})>0$, and, for any nonempty
word $f_{i_{1}}\cdots f_{i_{m}}\in B$, $\rho (f_{i_{1}}\cdots
f_{i_{m}})=\max \{\rho (f_{i_{j}}):j=1,\ldots ,m\}$.
\end{itemize}

\begin{definition}
A \emph{bounded} complexity measure is a complexity measure $\psi $, which has the \emph{%
boundedness from below} property: for any word $\alpha \in B$, $\psi (\alpha
)\geq \left\vert \alpha \right\vert $.
\end{definition}

Any complexity measure satisfying the equality (\ref{E1}), in particular the
function $h$, is a bounded complexity measure. One can show that if functions $%
\psi _{1}$ and $\psi _{2}$ are complexity measures, then the functions $\psi _{3}$
and $\psi _{4}$ are complexity measures, where for any $\alpha \in B$, $\psi
_{3}(\alpha )=\psi _{1}(\alpha )+\psi _{2}(\alpha )$ and $\psi _{4}(\alpha
)=\max (\psi _{1}(\alpha ),\psi _{2}(\alpha ))$. If the function $\psi _{1}$
is a bounded complexity measure, then the functions $\psi _{3}$ and $\psi _{4}$
are bounded complexity measures.

\begin{definition}
    Let $\psi $ be a complexity measure. We extend it to the set of all finite
subsets of the set $P$. Let $D$ be a finite subset of the set $P$. If $%
D=\emptyset $, then $\psi (D)=0$. Let $D=\{f_{i_{1}},\ldots ,f_{i_{m}}\}$
and $m\geq 1$. Then $\psi (D)=\psi (f_{i_{1}}\cdots f_{i_{m}})$.
\end{definition}

\subsection{Parameters of Decision Trees and Tables\label{S1.5}}

Let $\Gamma \in \mathcal{C}_{k}$ and $\tau =v_{1},d_{1},\ldots
,v_{m},d_{m},v_{m+1}$ be a complete path of $\Gamma $. We correspond to the
path $\tau $ a word $F(\tau )\in B$: if $m=1$, then $F(\tau )=\lambda $, and
if $m>1$ and, for $j=2,\ldots ,m$, the node $v_{j}$ is labeled with the
attribute $f_{i_{j}}$, then $F(\tau )=f_{i_{2}}\cdots f_{i_{m}}$.

\begin{definition}
    Let $\psi $ be a complexity measure. We extend the function $\psi $ to the set $%
\mathcal{C}_{k}$. Let $\Gamma \in \mathcal{C}_{k}$. Then $\psi (\Gamma
)=\max \{\psi (F(\tau ))\}$, where the maximum is taken over all complete
paths $\tau $ of the decision tree $\Gamma $. For a given complexity measure $%
\psi $, the value $\psi (\Gamma )$ will be called the \emph{complexity of the
decision tree} $\Gamma $. The value $h(\Gamma )$ will be called the \emph{%
depth} of the decision tree $\Gamma $.
\end{definition}

Let $\psi $ be a complexity measure. We now describe the functions $\psi ^{d}$, $%
\psi ^{s}$, $\Theta _{\psi }$, $W_{\psi }$, $V_{\psi }$, $S_{\psi }$, $\hat{S%
}_{\psi }$, $M_{\psi }$, and $N$ defined on the set $\mathcal{M}_{k}^{2}$
and taking values from the set $\omega $. By definition, the value of each
of these functions for  $\Lambda $ is equal $0$. Let $T\in \mathcal{M}%
_{k}^{2}\setminus \{\Lambda \}$.

\begin{itemize}
\item $\psi ^{d}(T)=\min \{\psi
(\Gamma )\}$, where the minimum is taken over all deterministic decision
trees $\Gamma $ for the table $T$.

\item If $T\in \mathcal{M}_{k}^{2}\mathcal{C}$, then $\psi ^{s}(T)=0$. If $%
T\notin \mathcal{M}_{k}^{2}\mathcal{C}$, then $\psi ^{s}(T)=\min \{\psi
(\Gamma )\}$, where the minimum is taken over all strongly nondeterministic
decision trees $\Gamma $ for the table $T$.

\item $\Theta _{\psi }(T)=\min \{\psi (D)\}$, where the minimum is taken
over all tests $D$ of the table $T$.

\item $W_{\psi }(T)=\psi (P(T))$.

\item $V_{\psi }(T)=\max \{\psi (f_{i}):f_{i}\in P(T)\}$.

\item Let $\bar{\delta}$ be a row of the table $T$. Denote $S_{\psi }(T,\bar{%
\delta})=\min \{\psi (D)\}$, where the minimum is taken over all subsets $D$
of the set $P(T)$ such that in the set of columns of $T$ labeled with
attributes from $D$ the row $\bar{\delta}$ is different from all other rows
of the table $T$. Then $S_{\psi }(T)=\max \{S_{\psi }(T,\bar{\delta})\}$,
where the maximum is taken over all rows $\bar{\delta}$ of the table $T$.

\item $\hat{S}_{\psi }(T)=\max \{S_{\psi }(T^{\ast }):T^{\ast }\in \left[ T%
\right] \}$.

\item If $T\in \mathcal{M}_{k}^{2}\mathcal{C}$, then $M_{\psi }(T)=0$. Let $%
T\notin \mathcal{M}_{k}^{2}\mathcal{C}$, $\left\vert P(T)\right\vert =n$,
and columns of the table $T$ be labeled with the attributes $%
f_{t_{1}},\ldots ,f_{t_{n}}$. Let $\bar{\delta}=(\delta _{1},\ldots ,\delta
_{n})\in E_{k}^{n}$. Denote by $M_{\psi }(T,\bar{\delta})$ the minimum
number $p\in \omega $ for which there exist attributes $f_{t_{i_{1}}},\ldots
,f_{t_{i_{m}}}\in P(T)$ such that $T(f_{t_{i_{1}}},\delta _{i_{1}})\cdots
(f_{t_{i_{m}}},\delta _{i_{m}})\in \mathcal{M}_{k}^{2}\mathcal{C}$ and $\psi
(f_{t_{i_{1}}}\cdots f_{t_{i_{m}}})=p$. Then $M_{\psi }(T)=\max \{M_{\psi
}(T,\bar{\delta}):\bar{\delta}\in E_{k}^{n}\}$.

\item $N(T)$ is the number of rows in the table $T$.
\end{itemize}

For the complexity measure $h$, we denote $\Theta (T)=\Theta _{h}(T)$, $W(T)=W_{h}(T)$%
, $S(T,\bar{\delta})=S_{h}(T,\bar{\delta})$, $S(T)=S_{h}(T)$, $\hat{S}(T)=%
\hat{S}_{h}(T)$, $M(T,\bar{\delta})=M_{h}(T,\bar{\delta})$, and $%
M(T)=M_{h}(T)$. Note that $W(T)$ is the number of columns in the table $T$.

\begin{example}
    We denote by $T_{0}$ the decision table shown in Fig. \ref{fig1}. One can show that $h^{d}(T_{0})=2$, $h^{s}(T_{0})=1$, $\Theta(T_{0})=2$, $W(T_{0})=3$, $N(T_{0})=6$, $S(T_{0})=2$, $\hat{S}(T_{0})=2$, and $M(T_{0})=2$.
\end{example}

\section{Main Results\label{S2}}

In this section, we consider results obtained for the functions $F_{\psi ,A}^{W}$, $F_{\psi ,A}^{\Theta}$, $F_{\psi ,A}$, and $G_{\psi ,A}$ and discuss closed classes of decision tables generated by information systems.

\subsection{Functions $F_{\protect\psi ,A}^{W}$ and $F_{\protect\psi %
,A}^{\Theta }$\label{S2.1}}

Let $\psi $ be a bounded complexity measure and $A$ be a nonempty closed class of
decision tables from $\mathcal{M}_{k}^{2}$. We now define functions $F_{\psi
,A}^{W}:\omega \rightarrow \omega $ and $F_{\psi ,A}^{\Theta }:\omega
\rightarrow \omega $. Let $n\in \omega $. Then%
\begin{eqnarray*}
F_{\psi ,A}^{W}(n) &=&\max \{\psi ^{d}(T):T\in A,W_{\psi }(T)\leq n\}, \\
F_{\psi ,A}^{\Theta }(n) &=&\max \{\psi ^{d}(T):T\in A,\Theta _{\psi
}(T)\leq n\}.
\end{eqnarray*}
The function $F_{\psi ,A}^{W}$ characterizes the growth in the worst case of the minimum complexity of a deterministic decision tree for a decision table with the growth of the complexity of the set of attributes attached to columns of the table.
The function $F_{\psi ,A}^{\Theta }$ characterizes the growth in the worst case of the minimum complexity of a deterministic decision tree for a decision table with the growth of the minimum complexity of a test of the table.

Let $D=\{n_{i}:i\in \omega \}$ be an infinite subset of the set $\omega $ in
which, for any $i\in \omega $, $n_{i}<n_{i+1}$. Let us define a function $%
H_{D}:\omega \rightarrow \omega $. Let $n\in \omega $. If $n<n_{0}$, then $%
H_{D}(n)=0$. If, for some $i\in \omega $, $n_{i}\leq n<n_{i+1}$, then $%
H_{D}(n)=n_{i}$.

\begin{theorem}
\label{T1}Let $\psi $ be a bounded complexity measure and $A$ be a nonempty
closed class of decision tables from $\mathcal{M}_{k}^{2}$. Then $F_{\psi
,A}^{W}$ and $F_{\psi ,A}^{\Theta }$ are nondecreasing functions for which
one of the following statements holds:

(a) If the functions $S_{\psi }$ and $N$ are bounded from above on the class
$A$, then there exists a nonnegative constant $c_{0}$ such that $F_{\psi
,A}^{W}(n)\leq F_{\psi ,A}^{\Theta }(n)\leq c_{0}$ for any $n\in \omega $.

(b) If the function $S_{\psi }$ is bounded from above on the class $A$ and
the function $N$ is not bounded from above on the class $A$, then $F_{\psi
,A}^{W}(0)=F_{\psi ,A}^{\Theta }(0)=0$ and there exist positive constants $%
c_{1}$, $c_{2}$, $c_{3}$, $c_{4}$ such that $c_{1}\log _{2}n-c_{2}\leq
F_{\psi ,A}^{W}(n)\leq F_{\psi ,A}^{\Theta }(n)\leq c_{3}\log _{2}n+c_{4}$
for any $n\in \omega \setminus \{0\}$.

(c) If the function $S_{\psi }$ is not bounded from above on the class $A$,
then there exists an infinite subset $D$ of the set $\omega $ such that $%
H_{D}(n)\leq F_{\psi ,A}^{W}(n)\leq F_{\psi ,A}^{\Theta }(n)\leq n$ for any $%
n\in \omega $.
\end{theorem}

For the complexity measure $h$, the bounds from the statement (c) of Theorem \ref%
{T1} can be improved.

\begin{theorem}
\label{T2}Let $A$ be a nonempty closed class of decision tables from $%
\mathcal{M}_{k}^{2}$ for which the function $S$ is not bounded from above.
Then $F_{h,A}^{W}(n)=F_{h,A}^{\Theta }(n)=n$ for any $n\in \omega $.
\end{theorem}

In the general case, the upper bound from the statement (c) of Theorem \ref%
{T1} is attained not for all $n\in \omega $.

\begin{theorem}
\label{T3}For any infinite subset $D$ of the set $\omega $ such that $0 \notin D$, there exist a
bounded complexity measure $\psi $ and a nonempty closed class $A$ of decision
tables from the set $\mathcal{M}_{k}^{2}$ for which $F_{\psi
,A}^{W}(n)=F_{\psi ,A}^{\Theta }(n)=H_{D}(n)$ for any $n\in \omega $.
\end{theorem}

\subsection{Function $F_{\protect\psi ,A}$\label{S2.2}}

Let $\psi $ be a bounded complexity measure and $A$ be a nonempty closed class of
decision tables from the set $\mathcal{M}_{k}^{2}$. We now define possibly
partial function $F_{\psi ,A}:\omega \rightarrow \omega $. Let $n\in \omega $%
. If the set $\{\psi ^{d}(T):T\in A,\psi ^{s}(T)\leq n\}$ is infinite, then
the value $F_{\psi ,A}(n)$ is undefined. Otherwise, $F_{\psi ,A}(n)=\max
\{\psi ^{d}(T):T\in A,\psi ^{s}(T)\leq n\}$.

The function $F_{\psi ,A}$ characterizes the growth in the worst case of the minimum complexity of a deterministic decision tree for a decision table with the growth of the minimum complexity of a strongly nondeterministic decision tree for the table.

Let us define possibly partial functions $S_{\psi ,A}:\omega \rightarrow
\omega $ and $N_{\psi ,A}:\omega \rightarrow \omega $. Let $n\in \omega $.
Denote $A_{\psi }(n)=\{T:T\in A,V_{\psi }(T)\leq n\}$. If the set $%
\{S(T):T\in A_{\psi }(n)\}$ is infinite, then the value $S_{\psi ,A}(n)$ is
undefined. Otherwise, $S_{\psi ,A}(n)=\max \{S(T):T\in A_{\psi }(n)\}$. If
the set $\{N(T):T\in A_{\psi }(n)\}$ is infinite, then the value $N_{\psi
,A}(n)$ is undefined. Otherwise, $N_{\psi ,A}(n)=\max \{N(T):T\in A_{\psi
}(n)\}$.

Let us indicate the condition for the function $F_{\psi ,A}$ to be defined
everywhere.

\begin{theorem}
\label{T4}Let $\psi $ be a bounded complexity measure and $A$ be a nonempty
closed class of decision tables from $\mathcal{M}_{k}^{2}$. Then the
function $F_{\psi ,A}$ is defined everywhere if and only if the function $%
N_{\psi ,A}$ is defined everywhere.
\end{theorem}

The following statement characterizes the possible behavior of the function $%
F_{\psi ,A}$.

\begin{theorem}
\label{T5} (a) Let $\psi $ be a bounded complexity measure, $A$ be a nonempty
closed class of decision tables from $\mathcal{M}_{k}^{2}$ and the function $%
F_{\psi ,A}$ be defined everywhere. Then $F_{\psi ,A}$ is a nondecreasing
function and either there exists a nonnegative constant $c$ for which $%
F_{\psi ,A}(n)\leq c$ for any $n\in \omega $ or there exists an infinite
subset $D$ of the set $\omega $ for which $F_{\psi ,A}(n)\geq H_{D}(n)$ for
any $n\in \omega $.

(b) Let $\varphi :\omega \rightarrow \omega $ be a nondecreasing function
for which $\varphi (n)\geq n$ for any $n\in \omega $ and $\varphi (0)=0$.
Then there exist a bounded complexity measure $\psi $ and a nonempty closed class
$A$ of decision tables from $\mathcal{M}_{k}^{2}$ for which, for any $n\in
\omega $, the value $F_{\psi ,A}(n)$ is defined and coincides with $\varphi
(n)$.
\end{theorem}

Of particular interest is the case when $F_{\psi ,A}$ is an everywhere
defined function bounded from above by some polynomial.

\begin{theorem}
\label{T6} Let $\psi $ be a bounded complexity measure, $A$ be a nonempty closed
class of decision tables from $\mathcal{M}_{k}^{2}$ and the function $%
F_{\psi ,A}$ be defined everywhere. Then a polynomial $p_{0}$ such that $%
F_{\psi ,A}(n)\leq p_{0}(n)$ for any $n\in \omega $ exists if an only if
there are exist polynomials $p_{1}$ and $p_{2}$ such that $S_{\psi
,A}(n)\leq p_{1}(n)$ and $N_{\psi ,A}(n)\leq 2^{p_{2}(n)}$ for any $n\in
\omega $.
\end{theorem}

\subsection{Function $G_{\protect\psi ,A}$\label{S2.1a}}

Let $\psi $ be a bounded complexity measure and $A$ be a nonempty closed
class of decision tables from $\mathcal{M}_{k}^{2}$. We now define a
function $G_{\psi ,A}$. Let $n\in \omega $. Then%
\begin{equation*}
G_{\psi ,A}(n)=\max \{\psi ^{s}(T):T\in A,W_{\psi }(T)\leq n\}.
\end{equation*}
The function $G_{\psi ,A}$ characterizes the growth in the worst case of the minimum complexity of a strongly nondeterministic decision tree for a decision table with the growth of the complexity of the set of attributes attached to columns of the table.

\begin{theorem}
\label{T1a}Let $\psi $ be a bounded complexity measure and $A$ be a nonempty
closed class of decision tables from $\mathcal{M}_{k}^{2}$. Then $G_{\psi ,A}
$ is a nondecreasing function for which one of the following statements
holds:

(a) If the function $S_{\psi }$ is bounded from above on the class $A$,
then there exists a nonnegative constant $c$ such that $G_{\psi ,A}(n)\leq c$
for any $n\in \omega $.

(b) If the function $S_{\psi }$ is not bounded from above on the class $A$,
then there exists an infinite subset $D$ of the set $\omega $ such that $%
H_{D}(n)\leq G_{\psi ,A}(n)\leq n$ for any $n\in \omega $.
\end{theorem}

For the complexity measure $h$, the bounds from the statement (b) of Theorem %
\ref{T1a} can be improved.

\begin{theorem}
\label{T2a}Let $A$ be a nonempty closed class of decision tables from $%
\mathcal{M}_{k}^{2}$ for which the function $S$ is not bounded from above.
Then $G_{h,A}(n)=n$ for any $n\in \omega $.
\end{theorem}

In the general case, the upper bound from the statement (b) of Theorem \ref%
{T1a} is attained not for all $n\in \omega $.

\begin{theorem}
\label{T3a}For any infinite subset $D$ of the set $\omega $ such that $0 \notin D$, there exist a
bounded complexity measure $\psi $ and a nonempty closed class $A$ of
decision tables from the set $\mathcal{M}_{k}^{2}$ for which $G_{\psi
,A}(n)=H_{D}(n)$ for any $n\in \omega $.
\end{theorem}

\subsection{Family of Closed Classes of Decision Tables\label{S2.3}}

Let $U$ be a set and $\Phi =\{f_{0},f_{1},\ldots \}$ be a finite or
countable set of functions (attributes) defined on $U$ and taking values
from $E_{k}$. The pair $(U,\Phi )$ is called a $k$-\emph{information system}%
. A $(2)$-\emph{problem} over $(U,\Phi )$ is an arbitrary tuple $z=(U,\nu
,f_{i_{1}},\ldots ,f_{i_{n}})$, where $n\in \omega \setminus \{0\}$, $\nu
:E_{k}^{n}\rightarrow E_{2}$ and $f_{i_{1}},\ldots ,f_{i_{n}}$ are functions from $\Phi $ with pairwise
different indices $i_1 , \ldots , i_n$. The problem $z$ is to determine the value $%
\nu (f_{i_{1}}(u),\ldots ,f_{i_{n}}(u))$ for a given $u\in U$. Various examples of $k$-information systems and problems over these systems including $2$-problems can be found in \cite{Moshkov05}.

We denote by $T(z)$ a decision table from $\mathcal{M}_{k}^{2}$ with $n$
columns labeled with attributes $f_{i_{1}},\ldots ,f_{i_{n}}$. A row $%
(\delta _{1},\ldots ,\delta _{n})\in E_{k}^{n}$ belongs to the table $T(z)$
if and only if the system of equations $\{f_{i_{1}}(x)=\delta _{1},\ldots
,f_{i_{n}}(x)=\delta _{n}\}$ has a solution from the set $U$. This row is
labeled with the decision $\nu (\delta _{1},\ldots ,\delta _{n})$.

Let the algorithms for the problem $z$ solving be algorithms in which each
elementary operation consists in calculating the value of some attribute
from the set $\{f_{i_{1}},\ldots ,f_{i_{n}}\}$ on the element $u\in U$.
Then, as a model of the problem $z$, we can use the decision table $T(z)$,
and as models of algorithms for the problem $z$ solving -- deterministic and
strongly nondeterministic decision trees for the table $T(z)$.

Denote by $Z(U,\Phi )$ the set of $(2)$-problems over $(U,\Phi )$ and $%
A(U,\Phi )=\{T(z):z\in Z(U,\Phi )\}$. One can show that $A(U,\Phi
)=[A(U,\Phi )]$, i.e., $A(U,\Phi )$ is a closed class of decision tables
from $\mathcal{M}_{k}^{2}$ \emph{generated} by the information system $(U,\Phi )$.

Closed classes of decision tables generated by $k$-information systems are the most
natural examples of closed classes. However, the notion of a closed class is
essentially wider. In particular, the union $A(U_1,\Phi_1 )\cup A(U_2,\Phi_2 )$, where $(U_1,\Phi_1 )$ and $(U_2,\Phi_2 )$ are $k$-information systems, is a closed class, but generally, we cannot find an
information system $(U,\Phi )$ such that $A(U,\Phi )= A(U_1,\Phi_1 )\cup A(U_2,\Phi_2 )$.

\subsection{Example of Information System\label{S2.4}}

Let $%
\mathbb{R}
$ be the set of real numbers and $F=\{f_{i}:i\in \omega \}$ be the set of
functions defined on $%
\mathbb{R}
$ and taking values from the set $E_{2}$ such that, for any $i\in \omega $
and $a\in
\mathbb{R}
$,%
\[
f_{i}(a)=\left\{
\begin{array}{ll}
0, & a<i, \\
1, & a\geq i.%
\end{array}%
\right.
\]

Let $\psi $ be a bounded complexity measure and $A=A(%
\mathbb{R}
,F)$. One can prove the following statements:

\begin{itemize}
\item The function $N$ is not bounded from above on the set $A$.

\item The function $S_{\psi }$ is bounded from above on the set $A$ if and
only if there exists a constant $c_{0}\geq 0$ such that $\psi (f_{i})\leq
c_{0}$ for any $i\in \omega $.

\item The function $N_{\psi ,A}$ is everywhere defined if and only if the
set $\{i:i\in \omega ,\psi (f_{i})\leq n\}$ is finite for any $n\in \omega $.

\item Polynomials $p_{1}$ and $p_{2}$ such that $S_{\psi ,A}(n)\leq p_{1}(n)$
and $N_{\psi ,A}(n)\leq 2^{p_{2}(n)}$ for any $n\in \omega $ exist if and
only if there exists a polynomial $p_{3}$ such that $\left\vert \{i:i\in
\omega ,\psi (f_{i})\leq n\}\right\vert \leq 2^{p_{3}(n)}$ for any $n\in
\omega $.
\end{itemize}

\section{Auxiliary Statements\label{S3}}

This section contains auxiliary statements. Some of them are of independent
interest.

\begin{lemma}
\label{L1} Let $T\in \mathcal{M}_{k}^{2}\setminus \{\Lambda \}$ and $\Gamma $
be a deterministic decision tree for the table $T$ or a strongly
nondeterministic decision tree for the table $T$. Then the set $P(\Gamma )$
is a test of the table $T$.
\end{lemma}

\begin{proof}
Let $T\in \mathcal{M}_{k}^{2}\mathcal{C}$. Then, by definition,  $P(\Gamma )$ is a test of the table $T$.

Let $T\notin \mathcal{M}_{k}^{2}\mathcal{C}$ and $\bar{\delta}_{0},\bar{%
\delta}_{1}$ be arbitrary rows of the table $T$ that are labeled with the
decisions $0$ and $1$, respectively. Since $\Gamma $ is a deterministic
decision tree for the table $T$ or a strongly nondeterministic decision tree
for the table $T$, then there exists a complete path $\tau $ of $\Gamma $
for which the row $\bar{\delta}_{1}$ belongs to the table $T(\tau )$ and all
rows of this table are labeled with the decision $1$. Evidently, $\bar{\delta%
}_{0}$ is not a row of the table $T(\tau )$. Therefore the rows $\bar{\delta}%
_{0}$ and $\bar{\delta}_{1}$ are different in a column labeled with an attribute $f_{i}$ such that $%
f_{i}\in P(\Gamma )$. Therefore $P(\Gamma )$ is a test of the table $T$.
\end{proof}

It is not difficult to prove the following statement.

\begin{lemma}
\label{L2} Let $T\in \mathcal{M}_{k}^{2}\setminus \{\Lambda \}$, $D$ be a
test of the table $T$ such that $D \neq \emptyset$, $T^{\ast }=I(P(T)\setminus D,T)$ and $\Gamma $ be a
deterministic decision tree for the table $T^{\ast }$. Then $\Gamma $ is a
deterministic decision tree for the table $T$.
\end{lemma}

The notions of a decision table and a deterministic decision tree used in
this paper are somewhat different from the corresponding notions used in \cite{Moshkov83}.
Taking into account these differences, it is easy to prove the following two
statements, which follow almost directly from Theorem 2.1, Lemma 1.3 and
Theorem 2.2 from \cite{Moshkov83}.

\begin{lemma}
\label{L3} For any complexity measure $\psi $ and any table $T$ from $\mathcal{M}%
_{k}^{2}$,
\[
\psi ^{d}(T)\geq M_{\psi }(T).
\]
\end{lemma}

\begin{lemma}
\label{L4} For any complexity measure $\psi $ and any table $T$ from $\mathcal{M}%
_{k}^{2}$,
\[
\psi ^{d}(T)\leq \left\{
\begin{array}{ll}
0, & M_{\psi }(T)=0, \\
M_{\psi }(T)\log _{2}N(T), & M_{\psi }(T)\geq 1.%
\end{array}%
\right.
\]
\end{lemma}

It is not difficult to prove the following statement.

\begin{lemma}
\label{L5}For any complexity measure $\psi $ and any table $T$ from $\mathcal{M}%
_{k}^{2}$,
\[
\psi ^{d}(T)\leq \Theta _{\psi }(T).
\]
\end{lemma}

\begin{lemma}
\label{L6}For any table $T\in \mathcal{M}_{k}^{2}\setminus \mathcal{M}%
_{k}^{2}\mathcal{C}$,%
\[
h^{d}(T)>\log _{k}\Theta (T).
\]
\end{lemma}

\begin{proof}
Let $\Gamma $ be a deterministic decision tree for the table $T$ such that $h(\Gamma )=h^{d}(T)$. By Lemma \ref{L1}%
, $P(\Gamma )$ is a test of the table $T$. Therefore $\Theta (T)\leq
\left\vert P(\Gamma )\right\vert $. Denote by $L_{w}(\Gamma )$ the number of
nodes of $\Gamma ,$ which are neither the root nor a terminal node.
Evidently, $\left\vert P(\Gamma )\right\vert \leq L_{w}(\Gamma )$. One can
show that $L_{w}(\Gamma )\leq \sum_{i=0}^{h(\Gamma )-1}k^{i}$. Therefore $%
\Theta (T)\leq \sum_{i=0}^{h(\Gamma )-1}k^{i}=\frac{k^{h(\Gamma )}-1}{k-1}%
<k^{h(\Gamma )}$. Since $T\notin \mathcal{M}_{k}^{2}\mathcal{C}$, $\Theta
(T)>0$. Hence $h(\Gamma )>\log _{k}\Theta (T)$. Taking into account that $%
h(\Gamma )=h^{d}(T)$, we obtain $h^d(T)>\log _{k}\Theta (T)$.
\end{proof}

\begin{lemma}
\label{L7}For any complexity measure $\psi $ and any table $T$ from $\mathcal{M}%
_{k}^{2}$,%
\[
M_{\psi }(T)\leq 2\hat{S}_{\psi }(T).
\]
\end{lemma}

\begin{proof}
Let $T\in \mathcal{M}_{k}^{2}\mathcal{C}$. Then $M_{\psi }(T)=0$. Therefore $%
M_{\psi }(T)\leq 2\hat{S}_{\psi }(T)$.

Let $T\notin \mathcal{M}_{k}^{2}\mathcal{C}$, $W(T)=n$ and $f_{t_{1}},\ldots
,f_{t_{n}}$ be attributes attached to columns of the table $T$. Denote $%
D=\{f_{i}:f_{i}\in P(T),\psi (f_{i})\leq S_{\psi }(T)\}$ and $T^{\ast
}=I(P(T)\setminus D,T)$. Evidently, $T^{\ast }\in \lbrack T]$. Taking into
account that the function $\psi $ has the nondecreasing property, we obtain
that any two rows of $T$ are different in columns labeled with attributes
from the set $D$. Let for the definiteness, $D=\{f_{t_{1}},\ldots
,f_{t_{m}}\}$.

Let $\bar{\delta}=(\delta _{1},\ldots ,\delta _{n})\in E_{k}^{n}$. Let $%
(\delta _{1},\ldots ,\delta _{m})$ be a row of $T^{\ast }$. Since $T^{\ast
}\in \lbrack T]$, there exist attributes $f_{t_{i_{1}}},\ldots
,f_{t_{i_{s}}} $ of the table $T^{\ast }$ such that the row $(\delta
_{1},\ldots ,\delta _{m})$ is different from all other rows of $T^{\ast }$
in columns labeled with  these attributes and $\psi (f_{t_{i_{1}}}\cdots f_{t_{i_{s}}})\leq \hat{S}%
_{\psi }(T)$. It is clear that $T(f_{t_{i_{1}}},\delta _{i_{1}})\cdots
(f_{t_{i_{s}}},\delta _{i_{s}})\in \mathcal{M}_{k}^{2}\mathcal{C}$.
Therefore $M_{\psi }(T,\bar{\delta})\leq \hat{S}_{\psi }(T)$.

Let $(\delta _{1},\ldots ,\delta _{m})$ be not a row of $T^{\ast }$. We
consider $m$ tables $T^{\ast }(f_{t_{1}},\delta _{1})$, $T^{\ast
}(f_{t_{1}},\delta _{1})(f_{t_{2}},\delta _{2})$, ..., $T^{\ast
}(f_{t_{1}},\delta _{1})\cdots (f_{t_{m}},\delta _{m})$. If $T^{\ast
}(f_{t_{1}},\delta _{1})=\Lambda $, then $T(f_{t_{1}},\delta _{1})=\Lambda $%
. Since $f_{t_{1}}\in D$, $\psi (f_{t_{1}})\leq S_{\psi }(T)$. Taking into
account that $S_{\psi }(T)\leq \hat{S}_{\psi }(T)$, we obtain $M_{\psi }(T,%
\bar{\delta})\leq \hat{S}_{\psi }(T)$. Let $T^{\ast }(f_{t_{1}},\delta
_{1})\neq \Lambda $. Then there exists $p\in \{1,\ldots ,m-1\}$ such that $%
T^{\ast }(f_{t_{1}},\delta _{1})\cdots (f_{t_{p}},\delta _{p})\neq \Lambda $
and $T^{\ast }(f_{t_{1}},\delta _{1})\cdots (f_{t_{p+1}},\delta
_{p+1})=\Lambda $. Denote $C=\{f_{t_{p+1}},f_{t_{p+2}},\ldots ,f_{t_{n}}\}$
and $T^{0}=I(C,T)$. Evidently, $T^{0}\in \lbrack T]$. Therefore in $T^{0}$
there are attributes $f_{t_{i_{1}}},\ldots ,f_{t_{i_{l}}}$ such that the row
$(\delta _{1},\ldots ,\delta _{p})$ is different from all other rows of the
table $T^{0}$ in columns labeled with these attributes and $\psi (f_{t_{i_{1}}}\cdots
f_{t_{i_{l}}})\leq \hat{S}_{\psi }(T)$. One can show that $$T^{\ast
}(f_{t_{i_{1}}},\delta _{i_{1}})\cdots (f_{t_{i_{l}}},\delta
_{i_{l}})=T^{\ast }(f_{t_{1}},\delta _{1})\cdots (f_{t_{p}},\delta _{p}).$$
Therefore $T^{\ast }(f_{t_{i_{1}}},\delta _{i_{1}})\cdots
(f_{t_{i_{l}}},\delta _{i_{l}})(f_{t_{p+1}},\delta _{p+1})=\Lambda $. Hence $$%
T(f_{t_{i_{1}}},\delta _{i_{1}})\cdots (f_{t_{i_{l}}},\delta
_{i_{l}})(f_{t_{p+1}},\delta _{p+1})=\Lambda .$$ Since $f_{t_{p+1}}\in D$, $%
\psi (f_{t_{p+1}})\leq S_{\psi }(T)\leq \hat{S}_{\psi }(T)$. Using
boundedness from above property of the function $\psi $, we obtain $\psi
(f_{t_{i_{1}}}\cdots f_{t_{i_{l}}}f_{t_{p+1}})\leq 2\hat{S}_{\psi }(T)$.
Therefore $M_{\psi }(T,\bar{\delta})\leq 2\hat{S}_{\psi }(T)$.

Thus, for any $\bar{\delta}\in E_{k}^{n}$, $M_{\psi }(T,\bar{\delta})\leq 2%
\hat{S}_{\psi }(T)$. As a result, we obtain $M_{\psi }(T)\leq 2\hat{S}_{\psi
}(T)$.
\end{proof}

It is not difficult to prove the following two upper bounds on the minimum
complexity of strongly nondeterministic decision trees for a table.

\begin{lemma}
\label{L8}For any complexity measure $\psi $ and any table $T$ from $\mathcal{M}%
_{k}^{2}$,%
\[
\psi ^{s}(T)\leq \psi ^{d}(T).
\]
\end{lemma}

\begin{lemma}
\label{L9}For any complexity measure $\psi $ and any table $T$ from $\mathcal{M}%
_{k}^{2}$,%
\[
\psi ^{s}(T)\leq S_{\psi }(T).
\]
\end{lemma}

\begin{lemma}
\label{L10}For any table $T$ from $\mathcal{M}_{k}^{2}\setminus \{\Lambda \}$%
,%
\[
N(T)\leq (kW(T))^{S(T)}.
\]
\end{lemma}

\begin{proof}
If $N(T)=1$, then $S(T)=0$ and the considered inequality holds. Let $N(T)>1$%
. Then $S(T)>0$. Denote $m=S(T)$. Evidently, for any row $\bar{\delta}$ of
the table $T$, there exist attributes $f_{i_{1}},\ldots ,f_{i_{m}}\in P(T)$
and numbers $\sigma _{1},\ldots ,\sigma _{m}\in E_{k}$ such that the table $%
T(f_{i_{1}},\sigma _{1})\cdots (f_{i_{m}},\sigma _{m})$ contains only the
row $\bar{\delta}$. Therefore there is a one-to-one mapping of rows of the
table $T$ onto some set $B$ of pairs of tuples of the kind $%
((f_{i_{1}},\ldots ,f_{i_{m}}),(\sigma _{1},\ldots ,\sigma _{m}))$ where $%
f_{i_{1}},\ldots ,f_{i_{m}}\in P(T)$ and $\sigma _{1},\ldots ,\sigma _{m}\in
E_{k}$. Evidently, $\left\vert B\right\vert \leq W(T)^{m}k^{m}$. Therefore $%
N(T)\leq (kW(T))^{S(T)}$.
\end{proof}

It is not difficult to prove the following statement by the induction on the
number of rows in the table.

\begin{lemma}
\label{L11}For any table $T$ from $\mathcal{M}_{k}^{2}\setminus \{\Lambda \}$%
,%
\[
\Theta (T)\leq N(T)-1.
\]
\end{lemma}

\begin{lemma}
\label{L12}For any complexity measure $\psi $ and any table $T$ from $\mathcal{M}%
_{k}^{2}$, which contains at least two rows, there exists a table $T^{\ast
}\in \lbrack T]$ such that $\psi^d (T^{\ast })=W_{\psi }(T^{\ast })=S_{\psi
}(T^{\ast })=S_{\psi }(T)$ and $\psi ^{s}(T^{\ast })\leq V_{\psi }(T)$.
\end{lemma}

\begin{proof}
Let $\bar{\delta}$ be a row of the table $T$ such that $S_{\psi }(T,\bar{%
\delta})=S_{\psi }(T)$. Let $D$ be a subset of the set $P(T)$ with the
minimum cardinality such that $\psi (D)=S_{\psi }(T,\bar{\delta})$ and in
the set of columns labeled with attributes from $D$ the row $\bar{\delta}$
is different from all other rows of the table $T$. Let $\bar{\sigma}$ be a
tuple obtained from the row $\bar{\delta}$ by the removal of all numbers
that are in the intersections with columns labeled with attributes from the
set $P(T)\setminus D$. Let $\nu :E_{k}^{\left\vert D\right\vert }\rightarrow
E_{2}$ and, for any $\bar{\gamma}\in E_{k}^{\left\vert D\right\vert }$, if $%
\bar{\gamma}=\bar{\sigma}$, then $\nu (\bar{\gamma})=0$ and if $\bar{\gamma}%
\neq \bar{\sigma}$, then $\nu (\bar{\gamma})=1$. Denote $T^{\ast }=J(\nu
,I(P(T)\setminus D,T))$. From the fact that $D$ has the minimum cardinality
and from the properties of the function $\psi $ it follows that, for any
attribute from the set $D$, there exists a row of $T$, which is different
from the row $\bar{\delta}$ only in the column labeled with the considered
attribute among attributes from $D$. Therefore, for any attribute of the
table $T^{\ast }$, there exists a row of $T^{\ast }$, which is different
from the row $\bar{\sigma}$ only in the column labeled with the considered
attribute. Thus,%
\begin{equation}
S_{\psi }(T^{\ast },\bar{\sigma})=W_{\psi }(T^{\ast }).  \label{E2}
\end{equation}%
Using properties of the function $\psi $, we obtain $S_{\psi }(T^{\ast
})\leq W_{\psi }(T^{\ast })$. From this inequality and from (\ref{E2}) it follows
that
\begin{equation}
S_{\psi }(T^{\ast })=W_{\psi }(T^{\ast }).  \label{E3}
\end{equation}%
One can show that $M_{\psi }(T^{\ast },\bar{\sigma})=S_{\psi }(T^{\ast },%
\bar{\sigma})$. Therefore $M_{\psi }(T^{\ast })\geq S_{\psi }(T^{\ast },\bar{%
\sigma})$. From this inequality and from Lemma \ref{L3} it follows that $%
\psi^d (T^{\ast })\geq S_{\psi }(T^{\ast },\bar{\sigma})$. Using (\ref{E2}),
we obtain $\psi^d (T^{\ast })\geq W_{\psi }(T^{\ast })$. Using Lemma \ref{L5} and
nondecreasing property of the function $\psi $, we obtain $\psi^d (T^{\ast
})\leq W_{\psi }(T^{\ast })$. Hence
\begin{equation}
\psi^d (T^{\ast })=W_{\psi }(T^{\ast }).  \label{E4}
\end{equation}%
By the choice of the set $D$, $W_{\psi }(T^{\ast })=S_{\psi }(T)$. From this
equality and from (\ref{E3}) and (\ref{E4}) it follows that $\psi^d (T^{\ast
})=W_{\psi }(T^{\ast })=S_{\psi }(T^{\ast })=S_{\psi }(T)$. One can show
that $\psi ^{s}(T^{\ast })\leq V_{\psi }(T)$.
\end{proof}

Let $G$ be an undirected graph whose nodes are colored in two colors. An
edge of the graph $G$ will be called a \emph{multi-colored} edge, if it is
incident to nodes having different colors. By induction on the number of
nodes of the graph, it is easy to prove the following statement.

\begin{lemma}
\label{L13a}Let $G$ be an arbitrary undirected graph without loops and
multiple edges. Then the nodes of the graph $G$ can be colored in two colors
such that at least half of the edges of $G$ will be multi-colored.
\end{lemma}

A table $T$ from $\mathcal{M}_{k}^{2}$ will be called \emph{critical} if $%
T\neq \Lambda $ and, for any attribute of $T$, there are two rows of the
table $T$ that differ only in the column labeled with this attribute.

\begin{lemma}
\label{L13}For any critical table $T$ from $\mathcal{M}_{k}^{2}$, there
exists a mapping $\nu :E_{k}^{W(T)}\rightarrow E_{2}$ such that%
\[
h^{d}(J(\nu ,T))>\log _{k}(W(T)/2).
\]
\end{lemma}

\begin{proof}
We correspond to any attribute $f_{i}\in P(T)$ two rows $\bar{\delta}%
_{0}(f_{i})$ and $\bar{\delta}_{1}(f_{i})$ of the table $T$, which differ
only in the column labeled with this attribute. Denote by $G$ the undirected
graph with the set of nodes $\{\bar{\delta}_{0}(f_{i}),\bar{\delta}%
_{1}(f_{i}):f_{i}\in P(T)\}$ and the set of edges $\{(\bar{\delta}%
_{0}(f_{i}),\bar{\delta}_{1}(f_{i})):f_{i}\in P(T)\}$. Using Lemma \ref{L13a}%
, we obtain that the nodes of $G$ can be colored in two colors (for example,
blue and green) such that at least $W(T)/2$ edges of $G$ will be
multi-colored. Let us define a mapping $\nu :E_{k}^{W(T)}\rightarrow E_{2}$.
Let $\bar{\delta}\in E_{k}^{W(T)}$. If $\bar{\delta}$ is a blue node of $G$,
then $\nu (\bar{\delta})=0$. Otherwise, $\nu (\bar{\delta})=1$. Denote $%
T^{\ast }=J(\nu ,T)$. Let $D$ be an arbitrary test of the table $T^{\ast }$.
Evidently, if $f_{i}\in P(T)$ and the edge $(\bar{\delta}_{0}(f_{i}),\bar{%
\delta}_{1}(f_{i}))$ is multi-colored, then $f_{i}\in D$. Therefore $\Theta
(T^{\ast })\geq W(T)/2$. Using Lemma \ref{L6}, we obtain $h^{d}(T^{\ast
})>\log _{k}\Theta (T^{\ast })\geq \log _{k}(W(T)/2)$.
\end{proof}

\begin{lemma}
\label{L14}For any table $T\in \mathcal{M}_{k}^{2}\setminus \{\Lambda \}$,
there exists a table $T^{\ast }\in \lbrack T]$ such that%
\[
h^{d}(T^{\ast })\geq \frac{\log _{k}N(T)}{\hat{S}(T)}-2.
\]
\end{lemma}

\begin{proof}
If $N(T)=1$, then, evidently, the considered statement holds. Let $N(T)\geq
2 $. Let $D$ be a subset of $P(T)$ with the minimum cardinality such that in
columns labeled with attributes from $D$ any two rows of $T$ are different.
Denote $T^{0}=I(P(T)\setminus D,T)$. One can show that $T^{0}$ is a critical
table. By Lemma \ref{L13}, there exists a mapping $\nu
:E_{k}^{W(T^{0})}\rightarrow E_{2}$ such that $h^{d}(J(\nu ,T^{0}))>\log
_{k}(W(T^{0})/2)\geq \log _{k}W(T^{0})-1$. Denote $T^{\ast }=J(\nu ,T^{0})$.
Evidently, $T^{\ast }\in \lbrack T]$. Taking into account that $%
W(T^{0})=W(T^{\ast })$, we obtain%
\begin{equation}
h^{d}(T^{\ast })>\log _{k}W(T^{\ast })-1.  \label{E5}
\end{equation}

Using Lemma \ref{L10}, we obtain $N(T)=N(T^{\ast })\leq (kW(T^{\ast
}))^{S(T^{\ast })}\leq (kW(T^{\ast }))^{\hat{S}(T)}$. Therefore $\hat{S}%
(T)(\log _{k}W(T^{\ast })+1)\geq \log _{k}N(T)$ and $\log _{k}W(T^{\ast
})\geq \frac{\log _{k}N(T)}{\hat{S}(T)}-1$. Using (\ref{E5}), we obtain $%
h^{d}(T^{\ast })\geq \frac{\log _{k}N(T)}{\hat{S}(T)}-2$.
\end{proof}

\section{Proofs of Theorems \protect\ref{T1}, \protect\ref{T2}, and \protect
\ref{T3}\label{S4}}

\begin{proof}[Proof of Theorem \ref{T1}] Evidently, $F_{\psi ,A}^{W}$ and $F_{\psi ,A}^{\Theta
} $ are nondecreasing functions. Using properties of commutativity and
nondecreasing of the function $\psi $, we obtain that, for any table $T$
from $\mathcal{M}_{k}^{2}$, $\Theta _{\psi }(T)\leq W_{\psi }(T)$.
Therefore, for any $n\in \omega $,
\begin{equation}
F_{\psi ,A}^{W}(n)\leq F_{\psi ,A}^{\Theta }(n).  \label{E6}
\end{equation}%
It is not difficult to show that
\begin{equation}
F_{\psi ,A}^{W}(0)=F_{\psi ,A}^{\Theta }(0).  \label{E7}
\end{equation}

(a) Let the functions $S_{\psi }$ and $N$ be bounded from above on the
class $A$. Then there are constants $a>0$ and $b>0$ such that $S_{\psi
}(T)\leq a$ and $N(T)\leq b$ for any table $T\in A$. Let $T\in A$. Taking
into account that $A$ is a closed class, we obtain $\hat{S}_{\psi }(T)\leq a$%
. By Lemma \ref{L7}, $M_{\psi }(T)\leq 2a$. From this inequality, inequality
$N(T)\leq b$ and from Lemma \ref{L4} it follows that $\psi ^{d}(T)\leq
2a\log _{2}b$. Denote $c_{0}=2a\log _{2}b$. Taking into account that $T$ is
an arbitrary table from the class $A$ and using (\ref{E6}), we obtain that,
for any $n\in \omega $,%
\[
F_{\psi ,A}^{W}(n)\leq F_{\psi ,A}^{\Theta }(n)\leq c_{0}.
\]

(b) Let the function $S_{\psi }$ be bounded from above on the closed class $%
A $ and the function $N$ be not bounded from above on $A$. Then there exists
a constant $a\geq 1$ such that, for any table $T\in A$,%
\begin{equation}
S_{\psi }(T)\leq a.  \label{E8}
\end{equation}%
Let $n\in \omega \setminus \{0\}$ and $T$ be an arbitrary table from $A$
such that $\Theta _{\psi }(T)\leq n$. If $\Theta _{\psi }(T)=0$, then using
Lemma \ref{L5}, we obtain
\begin{equation}
\psi ^{d}(T)=0.  \label{E9}
\end{equation}%
Let $\Theta _{\psi }(T)\neq 0$ and $D$ be a test of the table $T$ such that $%
\psi (D)=\Theta _{\psi }(T)$. Denote $T^{\ast }=I(P(T)\setminus D,T)$.
Evidently, $T^{\ast }\in A$ and $T^{\ast }\neq \Lambda $. Using (\ref{E8})
and the boundedness from below property of the function $\psi $, we obtain
\begin{equation}
S(T^{\ast })\leq a.  \label{E10}
\end{equation}%
It is clear that $W(T^{\ast })=\left\vert D\right\vert $. Using the
boundedness from below property of the function $\psi $, we obtain $%
\left\vert D\right\vert \leq \psi (D)=\Theta _{\psi }(T)$. Therefore $%
W(T^{\ast })\leq n$. From this inequality, inequality (\ref{E10}), relation $%
T^{\ast }\neq \Lambda $, and Lemma \ref{L10} it follows that $N(T^{\ast
})\leq (kn)^{a}$. From (\ref{E8}) and Lemma \ref{L7} it follows that $%
M_{\psi}(T^{\ast })\leq 2a$. Using the last two inequalities and Lemma \ref{L4}, we
obtain%
\begin{equation}
\psi ^{d}(T^{\ast })\leq 2a^{2}\log _{2}n+2a^{2}\log _{2}k.  \label{E11}
\end{equation}%
Denote $c_{3}=2a^{2}$ and $c_{4}=2a^{2}\log _{2}k$. From Lemma \ref{L2} it
follows that $\psi ^{d}(T)\leq \psi ^{d}(T^{\ast })$. From this inequality
and (\ref{E11}) it follows that $\psi ^{d}(T)\leq c_{3}\log _{2}n+c_{4}$.
Taking into account that $n$ is an arbitrary number from $\omega \setminus
\{0\}$ and $T$ is an arbitrary table from $A$ such that $0<\Theta _{\psi
}(T)\le n$ and using (\ref{E9}), we obtain that, for any $n\in \omega \setminus
\{0\}$,%
\begin{equation}
F_{\psi ,A}^{\Theta }(n)\leq c_{3}\log _{2}n+c_{4}.  \label{E12}
\end{equation}

Let $n\in \omega \setminus \{0\}$. Denote $c_{1}=1/\log _{2}k$ and $%
c_{2}=\log _{k}(4a)$. We now show that
\begin{equation}
F_{\psi ,A}^{W}(n)\geq c_{1}\log _{2}n-c_{2}.  \label{E13}
\end{equation}%
If $n\leq 4a$, then, evidently, the considered inequality holds. Let $n>4a$.
Denote $m=\left\lfloor n/a\right\rfloor $. Taking into account that the
function $N$ is not bounded from above on the set $A$, we obtain that there
exists a table $T\in A$ such that $N(T)\geq k^{m}$. Let $B$ be a subset of
the set $P(T)$ with the minimum cardinality such that in the columns labeled
with attributes from $B$ any two rows of the table $T$ are different and,
for any $f_{i}\in B$, $\psi (f_{i})\leq a$. The existence of such set
follows from the inequality (\ref{E8}) and properties of commutativity and
nondecreasing of the function $\psi $. Evidently, $\left\vert B\right\vert
\geq m$. Let $D$ be a subset of the set $B$ such that $\left\vert
D\right\vert =m$. Denote $T^{0}=I(P(T)\setminus D,T)$. One can show that $%
T^{0}$ is a critical table. Using Lemma \ref{L13}, we obtain that there
exists a mapping $\nu :E_{k}^{m}\rightarrow E_{2}$ such that $h^{d}(J(\nu
,T^{0}))>\log _{k}(m/2)=\log _{k}(\left\lfloor n/a\right\rfloor /2)$. Denote
$T^{\ast }=J(\nu ,T^{0})$. Taking into account that $n>4a$, we obtain $%
\left\lfloor n/a\right\rfloor /2\geq n/(4a)$. Therefore $h^{d}(T^{\ast
})>\log _{k}(n/(4a))$. Since $\psi $ is a bounded complexity measure, $\psi
^{d}(T^{\ast })>\log _{k}(n/(4a))$ and $W_{\psi }(T^{\ast })\leq
\left\lfloor n/a\right\rfloor a\leq n$. Hence the inequality (\ref{E13})
holds. From (\ref{E6}), (\ref{E7}), (\ref{E12}), and (\ref{E13}) it follows
that $F_{\psi ,A}^{W}(0)=F_{\psi ,A}^{\Theta }(0)=0$ and, for any $n\in
\omega \setminus \{0\}$,%
\[
c_{1}\log _{2}n-c_{2}\leq F_{\psi ,A}^{W}(n)\leq F_{\psi ,A}^{\Theta
}(n)\leq c_{3}\log _{2}n+c_{4}.
\]

(c) Let the function $S_{\psi }$ be not bounded from above on the class $A$.
Let $n\in \omega \setminus \{0\}$ and $T$ be a table from $A$ such that $%
\Theta _{\psi }(T)\leq n$. Using Lemma \ref{L5}, we obtain that $\psi
^{d}(T)\leq \Theta _{\psi }(T)$. Therefore%
\begin{equation}
F_{\psi ,A}^{\Theta }(n)\leq n.  \label{E14}
\end{equation}

Using Lemma \ref{L12}, we obtain that the set $D=\{W_{\psi }(T):T\in A,\psi
^{d}(T)=W_{\psi }(T)\}$ is infinite. Evidently, for any $n\in D$, $F_{\psi
,A}^{W}(n)\geq n$. Taking into account that $F_{\psi ,A}^{W}$ is a
nondecreasing function, we obtain that, for any $n\in \omega \setminus \{0\}$%
,
\[
F_{\psi ,A}^{W}(n)\geq H_{D}(n).
\]%
From this inequality and from (\ref{E6}), (\ref{E7}), and (\ref{E14}) it
follows that, for any $n\in \omega $, $H_{D}(n)\leq F_{\psi
,A}^{W}(n)\leq F_{\psi ,A}^{\Theta }(n)\leq n$.
\end{proof}

\begin{proof}[Proof of Theorem \ref{T2}] From Theorem \ref{T1} it follows that, for any $n\in
\omega $, $0 \le F_{h,A}^{W}(n)\leq F_{h,A}^{\Theta }(n)\leq n$. In particular, $F_{h,A}^{W}(0)=F_{h,A}^{\Theta }(0)=0$. Let $n\in \omega
\setminus \{0\}$. Since the function $S$ is not bounded from above on the
class $A$, there exists a table $T\in A$ such that $S(T)\geq n$. Let $\bar{%
\delta}$ be a row of the table $T$ for which $S(T,\bar{\delta})=S(T)$. Let $%
C $ be a set of attributes of $T$ with the minimum cardinality such that in
columns labeled with attributes from $C$ the row $\bar{\delta}$ is different
from all other rows of the table $T$. Evidently, $\left\vert C\right\vert
\geq n$. Let $C^{\ast }$ be a subset of the set $C$ with $\left\vert C^{\ast
}\right\vert =n$. Denote $%
T^{0}=I(P(T) \setminus C^{\ast },T)$. One can show that $S(T^{0})=n$. From Lemma \ref%
{L12} it follows that there exists a table $T^{\ast }\in \lbrack T^{0}]$ for
which $h^d(T^{\ast })=W(T^{\ast })=n$. Therefore $F_{h,A}^{W}(n)\geq n$.
\end{proof}

\begin{proof}[Proof of Theorem \ref{T3}] Let $D=\{n_{i}:i\in \omega \}$, $0 \notin D$ and, for any $i\in
\omega $, $n_{i}<n_{i+1}$. For $i\in \omega $, we denote by $T_{i}$ a
decision table with one column labeled with the attribute $f_{i}$ and two
rows $(0)$ and $(1)$ labeled with decisions $0$ and $1$, respectively.
Denote $A=[\{T_{i}:i\in D\}]$. We now define a complexity measure $%
\psi $. Let $\psi (\lambda )=0$, $\psi (f_{0})=1$, $\psi (f_{i})=i$ for any $i\in \omega
\setminus \{0\}$, and $\psi (f_{i_{1}}\cdots f_{i_{m}})=\sum_{j=1}^{m}\psi
(f_{i_{j}})$ for any nonempty word $f_{i_{1}}\cdots f_{i_{m}}\in B$.
Evidently, $\psi $ is a bounded complexity measure. Let $n<n_{0}$. Then $F_{\psi
,A}^{W}(n)=F_{\psi ,A}^{\Theta }(n)=0$. Let $n_{i}\leq n<n_{i+1}$. One can
show that in this case $F_{\psi ,A}^{W}(n)=F_{\psi ,A}^{\Theta }(n)=n_{i}$.
\end{proof}

\section{Proofs of Theorems \protect\ref{T4}, \protect\ref{T5}, and \protect
\ref{T6}\label{S5}}

\begin{proof}[Proof of Theorem \ref{T4}] (a) Let the function $N_{\psi ,A}$ be everywhere
defined. Let $n\in \omega $, $T\in A$ and $\psi ^{s}(T)\leq n$. Let $T\notin
\mathcal{M}_{k}^{2}\mathcal{C}$.  Let $\Gamma $ be a strongly
nondeterministic decision tree for the table $T$ such that $\psi
(\Gamma )=\psi ^{s}(T)$. Using Lemma \ref{L1}, we obtain that $P(\Gamma )$
is a test of the table $T$. Denote $T^{\ast }=I(P(T)\setminus P(\Gamma ),T)$%
. Using the properties of commutativity and nondecreasing of the function $%
\psi $, we obtain that, for any $f_{i}\in P(\Gamma )$, $\psi (f_{i})\leq
\psi (\Gamma )$ and hence
\begin{equation}
\psi (f_{i})\leq n.  \label{E15}
\end{equation}%
Therefore $T^{\ast }\in A_{\psi }(n)$. Using this relation and Lemma \ref%
{L11}, we obtain that $\Theta (T^{\ast })<N(T^{\ast })-1<N_{\psi ,A}(n)$.
From these inequalities and from the boundedness from above property of the
function $\psi $ it follows that
\begin{equation}
\Theta _{\psi }(T^{\ast })<nN_{\psi ,A}(n).  \label{E16}
\end{equation}%
Using Lemma \ref{L5}, we obtain that $\psi ^d (T^{\ast })\leq \Theta _{\psi
}(T^{\ast })$. Taking into account that $P(\Gamma )$ is a test of the table $T
$ and using Lemma \ref{L2}, we obtain that $\psi ^{d}(T)\leq \psi
^{d}(T^{\ast })$. From the last two inequalities and from (\ref{E16}) it
follows that $\psi ^{d}(T)<nN_{\psi ,A}(n)$. If $T\in \mathcal{M}_{k}^{2}%
\mathcal{C}$, then $\psi ^{d}(T)=0.$ Therefore the last inequality holds.
Hence the set $\{\psi ^{d}(T):T\in A,\psi ^{s}(T)\leq n\}$ is finite and the
value $F_{\psi ,A}(n)$ is defined.

(b) Let the function $N_{\psi ,A}$ be not everywhere defined and the
function $S_{\psi ,A}$ be everywhere defined. Then there exists $n\in \omega
$ such that the set $\Omega =\{N(T):T\in A_{\psi }(n)\}$ is infinite. Let $%
T\in A_{\psi }(n)\setminus \{\Lambda \}$. Taking into account that $%
[T]\subseteq A_{\psi }(n)$ and using Lemma \ref{L14}, we obtain that there
exists a table $T^{\ast }$ from $A_{\psi }(n)$ for which $h^{d}(T^{\ast
})\geq \frac{\log _{k}N(T)}{\hat{S}(T)}-2$.\ Using the boundedness from
below property of the function $\psi $, we obtain $\psi ^{d}(T^{\ast })\geq
h^{d}(T^{\ast })$. Evidently, $\hat{S}(T)\leq S_{\psi ,A}(n)$. Hence $\psi
^{d}(T^{\ast })\geq \frac{\log _{k}N(T)}{S_{\psi ,A}(n)}-2$. Taking into
account that the set $\Omega $ is infinite, we obtain that the set $\Delta
=\{\psi ^{d}(T):T\in A_{\psi }(n)\}$ is infinite. Using Lemma \ref{L9}, we
obtain that $\psi ^{s}(T)\leq nS_{\psi ,A}(n)$ for any table $T\in A_{\psi
}(n)$. From this inequality and from the fact that the set $\Delta $ is
infinite it follows that the set $\{\psi ^{d}(T):T\in A,\psi ^{s}(T)\leq
nS_{\psi ,A}(n)\}$ is infinite. Therefore the value $F_{\psi ,A}(nS_{\psi
,A}(n))$ is not defined.

(c) Let the functions $N_{\psi ,A}$ and $S_{\psi ,A}$ be not everywhere
defined. Then there exists $n\in \omega $ for which the set $\{S(T):T\in A_{\psi }(n)\}$ is infinite. Therefore the set $\{S_{\psi}(T):T\in A_{\psi }(n)\}$ is infinite.
Using Lemma \ref{L12}, we obtain that
the set $\{\psi ^{d}(T):T\in A,\psi ^{s}(T)\leq n\}$ is infinite. Hence, the
value $F_{\psi ,A}(n)$ is not defined.
\end{proof}

\begin{proof}[Proof of Theorem \ref{T5}] (a) From the definition of the function $F_{\psi ,A}$
it follows that it is nondecreasing. Let the function $\psi ^d$ be bounded
from above on the set $A$ by a constant $c$. Then, evidently, $F_{\psi
,A}(n)\leq c$ for any $n\in \omega $. Let the function $\psi ^d$ be not
bounded from above on the set $A$. Let us assume that the function $\psi ^{s}
$ is bounded from above on the set $A$ by a constant $d\in \omega $. Then
the value $F_{\psi ,A}(d)$ is not defined, which is impossible. Thus, the
set $D=\{\psi ^{s}(T):T\in A\}$ is infinite. Using Lemma \ref{L8}, we obtain
that $\psi ^{d}(T)\geq \psi ^{s}(T)$ for any $T\in A$. Therefore $F_{\psi
,A}(n)\geq n$ for any $n\in D$. Taking into account that $F_{\psi ,A}$ is a
nondecreasing function, we obtain that $F_{\psi ,A}(n)\geq H_{D}(n)$ for any
$n\in \omega $.

\begin{figure}
\begin{minipage}[c]{1.0\textwidth}
\begin{center}
\begin{tabular}{ |ccccc|c| }
 \hline
 $f_{t(n)+1}$ & $f_{t(n)+2}$ & $f_{t(n)+3}$ & $\cdots$ & $f_{t(n)+\lceil\frac{\varphi(n)}{n}\rceil}$ &\\
  \hline
 0 & 0 & 0 & $\cdots$ & 0 & $0$ \\
 1 & 0 & 0 & $\cdots$ & 0 & $1$ \\
 0 & 1 & 0 & $\cdots$ & 0 & $1$ \\
 0 & 0 & 1 & $\cdots$ & 0 & $1$ \\
 $\vdots$ & $\vdots$ & $\vdots$ & $\vdots$ & $\vdots$ & $\vdots$\\
  0 & 0 & 0 & $\cdots$ & 1 & $1$ \\
 \hline
\end{tabular}
\end{center}
\end{minipage}
\caption{Decision table $T_{n}$}
\label{fig5}
\end{figure}

(b) Let $n\in \omega \setminus \{0\}$. Denote by $T_{n}$ the decision table
depicted in Fig. \ref{fig5}, where $t(n)=\sum_{i=1}^{n-1}\left\lceil \frac{\varphi (i)%
}{i}\right\rceil $ if $n>1$ and $t(1)=0$. Denote $A=[\{T_{n}:n\in \omega
\setminus \{0\}\}]$. We now define a bounded complexity measure $\psi $. Let $%
i\in \omega $. If $i=0$, then $\psi (f_{i})=1$. Let for some $n\in \omega
\setminus \{0\}$, $t(n)<i<t(n+1)$. Then $\psi (f_{i})=n$. Let, for some $%
n\in \omega \setminus \{0\}$, $i=t(n+1)$ and $\varphi (n)=ln+j$, where  $l \in \omega \setminus \{0\}$ and $%
0\leq j<n$. If $j=0$, then $\psi (f_{i})=n$. If $j>0$, then $\psi (f_{i})=j$%
. For any nonempty word $f_{i_{1}}\cdots f_{i_{m}}\in B$, $\psi
(f_{i_{1}}\cdots f_{i_{m}})=\sum_{j=1}^{m}\psi (f_{i_{j}})$. It is easy to show that $W_{\psi}(T_n)=\varphi (n)$ for any $n \in \omega \setminus \{0\}$.

One can show that the function $N_{\psi ,A}$ is everywhere defined. Using
Theorem \ref{T4}, we obtain that the function $F_{\psi ,A}$ is everywhere
defined. Evidently, $F_{\psi ,A}(0)=0=\varphi (0)$. Let $n\in \omega \setminus \{0\}$.
One can show that $\psi ^{s}(T_{n})=n$ and $M_{\psi }(T_{n},\bar{\delta}%
)=\varphi (n)$, where $\bar{\delta}$ is the first row of the table $T_{n}$.
Using Lemma \ref{L3}, we obtain $\psi^d (T_{n})\geq \varphi (n)$. Hence%
\begin{equation}
F_{\psi ,A}(n)\geq \varphi (n).  \label{E17}
\end{equation}

We now show that, for any $n\in \omega \setminus \{0\}$,%
\begin{equation}
F_{\psi ,A}(n)\leq \varphi (n).  \label{E18}
\end{equation}%
 Let $n\in \omega \setminus \{0\}$%
, $T\in A$ and $\psi ^{s}(T)\leq n$. Let $T\in \lbrack \{T_{1},\ldots
,T_{n}\}]$. Using Lemma \ref{L5}, we obtain that $\psi ^{d}(T)\leq \Theta
_{\psi }(T)$. From here and from the properties of commutativity and
nondecreasing of the function $\psi $ it follows that $\psi ^{d}(T)\leq
W_{\psi }(T)$. One can show that $W_{\psi }(T)\le \max \{W_{\psi
}(T_{1}),\ldots ,W_{\psi }(T_{n})\}\leq \varphi (n)$. Hence $\psi
^{d}(T)\leq \varphi (n)$.

Let $T\in \lbrack \{T_{n+1},T_{n+2},\ldots \}]$. Let $\Gamma $ be a strongly
nondeterministic decision tree for the table $T$ such that $\psi (\Gamma
)\leq n$. Using properties of commutativity and nondecreasing of the
function $\psi $, we obtain that $\psi (f_{i})\leq n$ for any $f_{i}\in
P(\Gamma )$. Evidently, $\left\vert \{f_{i}:f_{i}\in P(T),\psi (f_{i})\leq
n\}\right\vert \leq 1$. Using the property of boundedness from above of the
function $\psi $, we obtain that $\psi (P(\Gamma ))\leq n$. From Lemma \ref%
{L1} it follows that $P(\Gamma )$ is a test of the table $T$. Therefore $%
\Theta _{\psi }(T)\leq n$. Using Lemma \ref{L5}, we obtain that $\psi
^{d}(T)\leq n$. Since $\varphi (n)\geq n$, $\psi ^{d}(T)\leq \varphi (n)$.
Thus, the inequality (\ref{E18}) holds. From (\ref{E17}) and (\ref{E18}) it
follows that $F_{\psi ,A}(n)=\varphi (n)$.
\end{proof}

\begin{proof}[Proof of Theorem \ref{T6}] (a) Let there exist polynomials $p_{1}$ and $p_{2}$
such that, for any $n\in \omega $, $S_{\psi ,A}(n)\leq p_{1}(n)$ and $%
N_{\psi ,A}(n)\leq 2^{p_{2}(n)}$. Without loss of generality, we can assume
that, for any $n\in \omega $, $p_{1}(n)\geq 0$ and $p_{2}(n)\geq 0$. Let $%
n\in \omega ,$ $T\in A$ and $\psi ^{s}(T)\leq n$. Let us show that
\begin{equation}
\psi ^{d}(T)\leq 2np_{1}(n)p_{2}(n).  \label{E19}
\end{equation}

Let $T\in \mathcal{M}_{k}^{2}\mathcal{C}$. Then, as it is easy to check, $%
\psi ^{d}(T)=0$. Therefore the inequality (\ref{E19}) holds. Let $T\notin
\mathcal{M}_{k}^{2}\mathcal{C}$ and $\Gamma $ be a strongly nondeterministic
decision tree for the table $T$ such that $\psi (\Gamma )\leq n$. Using
Lemma \ref{L1}, we obtain that $P(\Gamma )$ is a test of the table $T$.
Taking into account the properties of commutativity and nondecreasing of the
function $\psi $, we obtain that, for any $f_{i}\in P(\Gamma )$, $\psi
(f_{i})\leq n$. Denote $T^{\ast }=I(P(T)\setminus P(\Gamma ),T)$. Evidently,
$T^{\ast }\in A_{\psi }(n)$. Using Lemma \ref{L7}, we obtain that $M(T^{\ast })\leq 2\hat{S}(T^{\ast })\leq 2S_{\psi ,A}(n)\leq
2p_{1}(n)$. Using the boundedness from above property of the function $\psi$, we obtain that $M_{\psi}(T^{\ast })\leq nM(T^{\ast })\leq 2np_1(n)$.
It is clear that $N(T^{\ast })\leq N_{\psi ,A}(n)\leq
2^{p_{2}(n)}$. From these inequalities and from Lemma \ref{L4} it follows
that $\psi^d (T^{\ast })\leq 2np_{1}(n)p_{2}(n)$. Using Lemma \ref{L2}, we
obtain that $\psi ^{d}(T)\leq \psi ^{d}(T^{\ast })$. Hence the inequality (%
\ref{E19}) holds. Taking into account that $T$ is an arbitrary table from $A$
for which $\psi ^{s}(T)\leq n$, we obtain $F_{\psi ,A}(n)\leq
2np_{1}(n)p_{2}(n)$.

(b) Let there be no a polynomial $p_{1}$ such that $S_{\psi ,A}(n)\leq
p_{1}(n)$ for any $n\in \omega $. Since the function $F_{\psi ,A}$ is
everywhere defined, from Theorem \ref{T4} it follows that $N_{\psi ,A}$ is
an everywhere defined function. Let $n\in \omega $ and $T\in A_{\psi }(n)$.
Using Lemma \ref{L11}, one can show that $S(T)\leq N(T)-1\leq N_{\psi ,A}(n)-1$.
Therefore the value $S_{\psi ,A}(n)$ is defined.
We now show that $F_{\psi ,A}(n) \ge S_{\psi ,A}(n)$. If $S_{\psi ,A}(n)=0$, then the considered inequality holds. Let $S_{\psi ,A}(n)>0$ and
$T^0$
be a table from $A_{\psi }(n)$ for which $S(T^0)=S_{\psi ,A}(n)$. From
Lemma \ref{L12} it follows that there exists a table $T^{\ast }\in [T^0]
$ for which $\psi ^{d}(T^{\ast })=S_{\psi }(T^0)$ and $\psi ^{s}(T^{\ast
})\leq V_{\psi }(T^0)$. Taking into account that $T\in A_{\psi }(n)$, we
obtain $\psi ^{s}(T^{\ast }) \le n$.
Using the boundedness from below property of  the function $\psi$, we obtain that $\psi ^{d}(T^{\ast })= S_{\psi}(T^0)\ge S(T^0)=S_{\psi ,A}(n)$.
Therefore $F_{\psi ,A}(n)\geq S_{\psi
,A}(n)$. Thus, there is no a polynomial $p_{0}$ such that $F_{\psi
,A}(n)\leq p_{0}(n)$ for any $n\in \omega $.

(c) Let there exist a polynomial $p_{1}$ such that $S_{\psi ,A}(n)\leq
p_{1}(n)$ for any $n\in \omega $ and there be no a polynomial $p_{2}$ such
that $N_{\psi ,A}(n)\leq 2^{p_{2}(n)}$ for any $n\in \omega $. Taking into
account that $F_{\psi ,A}$ is an everywhere defined function and using
Theorem \ref{T4} we obtain that $N_{\psi ,A}$ is everywhere defined. Let $%
n\in \omega $ and $T$ be a table from $A_{\psi }(n)$ for which $N(T)=N_{\psi
,A}(n)$.
Let $N_{\psi,A}(n) > 0$.
Using Lemma \ref{L14}, we obtain that there exists a table $T^{\ast
}\in A_{\psi }(n)$ for which $h^d(T^{\ast })\geq \frac{\log _{k}N(T)-2}{\hat{S}%
(T)}$. Then $\log _{k}N(T)\leq h^d(T^{\ast })\hat{S}(T)+2$. Taking into
account that $\psi $ is a bounded complexity measure, we obtain that $\hat{S}%
(T)\leq S_{\psi ,A}(n)\leq p_{1}(n)$ and $h^{d}(T^{\ast })\leq \psi
^{d}(T^{\ast })$. Using Lemma \ref{L9}, we obtain that $\psi ^{s}(T^{\ast
})\leq S_{\psi }(T^{\ast })\leq nS_{\psi ,A}(n)\leq np_{1}(n)$. Therefore $%
\psi ^{d}(T^{\ast })\leq F_{\psi ,A}(np_{1}(n))$. Hence $\log _{k}N_{\psi
,A}(n)\leq F_{\psi ,A}(np_{1}(n))p_{1}(n)+2$. Let us assume that there exists
a polynomial $p_{0}$ such that $F_{\psi ,A}(n)\leq p_{0}(n)$ for any $n\in
\omega $. Then, for any $n\in \omega $, $N_{\psi ,A}(n)\leq
2^{(p_{0}(np_{1}(n))p_{1}(n)+2)\log _{2}k}$, which is impossible. Thus, there
is no a polynomial $p_{0}$ such that $F_{\psi ,A}(n)\leq p_{0}(n)$ for any $%
n\in \omega $.
\end{proof}

\section{Proofs of Theorems \protect\ref{T1a}, \protect\ref{T2a}, and
\protect\ref{T3a}\label{S6}}

\begin{proof}[Proof of Theorem \ref{T1a}] Evidently, $G_{\psi ,A}$ is a nondecreasing function and $G_{\psi ,A}(0)=0$.

(a) Let the function $S_{\psi }$ be bounded from above on the class $A$.
Then there is a constant $c\ge 0$ such that $S_{\psi }(T)\leq c$ for any $T\in A$.
By Lemma \ref{L9}, $\psi ^{s}(T)\leq S_{\psi }(T)\leq c$ for any $T\in A$. Therefore,
for any $n\in \omega $,
$
G_{\psi ,A}(n)\leq c
$.

(b) Let the function $S_{\psi }$ be not bounded from above on the class $A$.
Then the set  $D=\{S_{\psi }(T):T\in A\}$ is infinite. We now show that $%
G_{\psi ,A}(n)\geq H_{D}(n)$ for any $n\in \omega $. Let $m\in D$, $T\in A$,
$S_{\psi }(T)=m$ and $\bar{\delta}$ be a row of $T$ such that $S_{\psi }(T,%
\bar{\delta})=m$. Let $Q$ be a subset of the set $P(T)$ with the minimum
cardinality such that $\psi (Q)=m$ and in columns labeled with the attributes
from the set $Q$ the row $\bar{\delta}$ is different from all other rows of
the table $T$. Denote $T^{0}=I(P(T)\setminus Q,T)$ and $\bar{\sigma}$ the
tuple obtained from $\bar{\delta}$ by the removal of numbers in
the intersection with columns labeled with the attributes from $P(T)\setminus Q$%
. One can show that $\bar{\sigma}$ is a row of $T^{0}$ and $S_{\psi }(T^0,\bar{%
\sigma})=m$. Let a mapping $\nu :E_{k}^{\left\vert Q\right\vert }\rightarrow
E_{2}$ be defined in the following way: for any $\bar{\gamma}\in
E_{k}^{\left\vert Q\right\vert }$, $\nu (\bar{\gamma})=1$ if and only if $%
\bar{\gamma}=\bar{\sigma}$. Denote $T^{\ast }=J(\nu ,T^{0})$. One can show
that $\psi ^{s}(T^{\ast })=W_{\psi }(T^{\ast })=m$. Therefore $G_{\psi
,A}(m)\geq m$. Taking into account that  $m$ is an arbitrary number from $D$ and $G_{\psi ,A}$ is a nondecreasing
function, we obtain that $G_{\psi ,A}(n)\geq H_{D}(n)$ for any $n\in \omega $%
. Let $T\in A$. By Lemma \ref{L9}, $\psi ^{s}(T)\leq S_{\psi }(T)$. It is
clear that $S_{\psi }(T)\leq W_{\psi }(T)$. Therefore $G_{\psi ,A}(n)\leq n$
for any $n\in \omega $.
\end{proof}

\begin{proof}[Proof of Theorem \ref{T2a}] From Theorem \ref{T1a} it follows that, for any $n\in
\omega $, $0\leq G_{h,A}(n)\leq n$. In particular, $G_{h,A}(0)=0$. Let $n\in
\omega \setminus \{0\}$. Since the function $S$ is not bounded from above on
the class $A$, there exists a table $T\in A$ such that $S(T)\geq n$. Let $%
\bar{\delta}$ be a row of the table $T$ such that $S(T,\bar{\delta})=S(T)$.
Let $C$ be a set of attributes of $T$ with the minimum cardinality such that
in columns labeled with the attributes from $C$ the row $\bar{\delta}$ is
different from all other rows of the table $T$. Evidently, $\left\vert
C\right\vert \geq n$. Let $C^{\ast }$ be a subset of the set $C$ with $%
\left\vert C^{\ast }\right\vert =n$. Denote $T^{0}=I(P(T)\setminus C^{\ast
},T)$. One can show that $S(T^{0})=n$. From Lemma \ref{L12} it follows that
there exists a table $T^{1}\in \lbrack T^{0}]$ for which $S(T^{1})=W(T^{1})=n
$. Let $\bar{\sigma}$ be a row of the table $T^{1}$ such that $S(T^{1},\bar{%
\sigma})=S(T^{1})$. Let a mapping $\nu :E_{k}^{W(T^{1})}\rightarrow E_{2}$
be defined in the following way: for any $\bar{\gamma}\in E_{k}^{W(T^{1})}$,
$\nu (\bar{\gamma})=1$ if and only if $\bar{\gamma}=\bar{\sigma}$. Denote $%
T^{\ast }=J(\nu ,T^{1})$. One can show that $\psi ^{s}(T^{\ast })=W(T^{\ast })=n$. Therefore $G_{h ,A}(n)\geq n$.
\end{proof}

\begin{proof}[Proof of Theorem \ref{T3a}] Let $D=\{n_{i}:i\in \omega \}$, $0\notin D$ and, for
any $i\in \omega $, $n_{i}<n_{i+1}$. For $i\in \omega $, we denote by $T_{i}$
a decision table with one column labeled with the attribute $f_{i}$ and two
rows $(0)$ and $(1)$ labeled with decisions $0$ and $1$, respectively.
Denote $A=[\{T_{i}:i\in D\}]$. We now define a complexity
measure $\psi $. Let $\psi (f_{0})=1$, $\psi (f_{i})=i$ for any $i\in \omega
\setminus \{0\}$, and $\psi (f_{i_{1}}\cdots f_{i_{m}})=\sum_{j=1}^{m}\psi
(f_{i_{j}})$ for any nonempty word $f_{i_{1}}\cdots f_{i_{m}}\in B$.
Evidently, $\psi $ is a bounded complexity measure. Let $n<n_{0}$. Then $%
G_{\psi ,A}(n)=0$. Let $n_{i}\leq n<n_{i+1}$. One can show that in this case
$G_{\psi ,A}(n)=n_{i}$.
\end{proof}

\section{Conclusions\label{S7}}

In this paper, we studied relationships among four parameters of tables from closed classes of decision tables with 0-1-decisions: the minimum complexity of a deterministic decision tree, the minimum complexity of a strongly nondeterministic decision tree, the complexity of the set of attributes attached to columns, and the minimum complexity of a test. Future research will be devoted to the similar study of deterministic and nondeterministic decision trees for tables from closed classes of conventional decision tables in which rows can be labeled with arbitrary decisions.

\subsection*{Acknowledgements}

Research reported in this publication was supported by King Abdullah
University of Science and Technology (KAUST).

\bibliographystyle{spmpsci}
\bibliography{closed_classes}

\end{document}